\documentstyle[aps, eqsecnum]{revtex}

\begin{document}

\draft

\title{Changes in Floquet State Structure at Avoided Crossings:
Delocalization and Harmonic Generation}
\author{T. Timberlake and L.~E. Reichl}
\address{Center for Studies in Statistical Mechanics and Complex
Systems\\
The University of Texas at Austin\\
Austin, Texas 78712}
\date{\today}
\maketitle

\begin{abstract}

Avoided crossings are common in the quasienergy spectra of strongly 
driven nonlinear quantum wells.  In this paper we examine the 
sinusoidally driven particle in a square potential well to show that 
avoided crossings can alter the structure of Floquet states in this 
system.  Two types of avoided crossings are identified: one type leads 
only to temporary changes (as a function of driving field strength) in 
Floquet state structure while the second type can lead to permanent 
delocalization of the Floquet states.  Radiation spectra from these 
latter states show a significant increase in high harmonic generation 
as the system passes through the avoided crossing.

\end{abstract}


\section{Introduction}
\label{intro}

Quantum systems whose classical counterparts are chaotic have received
much attention in recent years \cite{reichl1}.
One quantum signature of the classical transition to chaos is the appearance
of avoided crossings in the spectrum of the
quantum system.
 Avoided crossings
occur when the curves of two quantum eigenvalues, as a function of
some nonlinearity parameter, come near to crossing but then repel each
other \cite{wigner}.  As the classical system becomes increasingly chaotic, the
number of avoided crossings in the quantum spectrum increases
\cite{delande2,wang,yang}.  This
repulsion of eigenvalue curves leads to a change from Poisson to
random-matrix statistics in the eigenvalue spacings.  These avoided
crossings also play an important role in dynamical tunneling
\cite{latka1,latka2}.

Another common feature of these classically chaotic quantum systems is 
that their eigenstates often show nonclassical localization.  The 
probability density for these localized states remains in a small 
region of the phase space, even though there is no classical barrier 
to prevent them from spreading.  Even systems with globally chaotic 
classical motion can have localized quantum eigenstates.  This 
phenomenon was originally termed scarring \cite{heller} and it was 
found that the localization tends to occur near the paths of classical 
periodic orbits.

Understanding localization is particularly important in light of some 
recent discoveries in the field of atomic physics.  Experiments on 
atoms illuminated by an intense laser field show that the atoms can 
emit photons whose frequencies are many times the frequency of the 
incident laser \cite{macklin,chang}.  Recent theoretical work has 
shown that this high harmonic generation is related to delocalization 
and chaos \cite{chism,averbukh2}.  Theoretical work has also shown 
that in some cases the ionization rate of these atoms may actually 
decrease as the intensity of the laser is increased 
\cite{pont,potvliege}.  Some experimental evidence for this 
stabilization has been found \cite{deboer}.  Localization plays a 
major role in stabilizing these atoms against ionization 
\cite{sundaram1}.  Therefore we would like to have a better 
understanding of how localized eigenstates are created and destroyed 
in these systems, and what relationship this has to the other 
phenomena of ``chaotic'' quantum systems.

In this paper we will show that avoided crossings have a major impact
on localization.  A connection between level repulsion and the
creation of extended (delocalized) states has already been found
\cite{lin2,delande1}.  However, these studies looked at statistical
properties of the system as a whole.  Here we will concentrate on the
changes in the structure of the quantum eigenstates at a single
avoided crossing.  We identify two distinct types of avoided crossings
which have different effects on eigenstate structure.  One type of
crossing produces significant changes in the eigenstates only for the
parameter values at which the avoided crossing actually occurs.  The
second type results in structural changes that persist for parameter
values that are beyond the avoided crossing.  Finally, we will
investigate what impact these structural changes have on the radiation
spectrum of the system.

\section{Classical and Quantum Dynamics of The Driven Square Well}
\label{model}

The model we will use to study these phenomena is the sinusoidally
driven square well.  We choose this model because of its simplicity
and its connection with experimental work in solid-state physics.
The driven square well serves as a highly simplified model for
experiments involving electrons confined in GaAs/AlGaAs wells and
subjected to intense far-infrared radiation
\cite{sherwin1,sherwin2}.  This model is also advantageous because it
has been the subject of many theoretical studies, both classical and
quantum.  Its basic properties are well understood, allowing us to
focus on the particular phenomenon we are interested in.

\subsection{Classical Dynamics}

The Hamiltonian for the driven square well is:
\begin{equation}
\tilde{H} = \frac{\tilde{p}^{2}}{2m} + \tilde{\epsilon} \tilde{x}
\cos{\tilde{\omega_{0}} \tilde{t}}, \ \left| \tilde{x} \right| \leq a,
\end{equation}
where $m$ is the mass, $\tilde{p}$ is the momentum, and $\tilde{x}$ is 
the position of the particle.  The width of the square well is $2a$.  
The driving field has amplitude $\tilde{\epsilon}$ and frequency 
$\tilde{\omega_{0}}$, with $\tilde{t}$ as the time coordinate.  This 
Hamiltonian can be made dimensionless using the scaling transformation 
introduced in \cite{lin3}, where $\tilde{H}=Hc$, $\tilde{x}=xa$, 
$\tilde{p}=p\sqrt{2mc}$, $\tilde{\epsilon}=\epsilon c/a$, 
$\tilde{t}=ta\sqrt{2m/c}$, and 
$\tilde{\omega_{0}}=(\omega_{0}/a)\sqrt{c/(2m)}$.  The 
parameter $c$ is a new unit of energy in terms of which the 
scaled Hamiltonian will be expressed.  The scaled Hamiltonian (in 
units of $c$) is:
\begin{equation}
H = p^{2} + \epsilon x \cos{\omega_{0} t}, \ \left| x \right| \leq 1
\end{equation}
where all quantities are now dimensionless.

Note that $\epsilon$ and $\omega_{0}$ are not independent parameters,
since the transformation $(\omega_{0},\epsilon) \rightarrow
(\omega_{0}\sqrt{c},\epsilon c)$ produces the same dynamics (with a
rescaling of the energy unit $c$).  Because of this scaling
law we can choose an arbitrary $\omega_{0}$, study the dynamics as a
function of $\epsilon$, and effectively analyze the dynamics for any
set of $(\omega_{0},\epsilon)$.  In this paper we choose
$\omega_{0}=80$.

The driving field induces a series of nonlinear resonances of odd 
order in the square well system, with higher order resonances occurring 
at lower energies \cite{lin3,fuka}.  This means that the $N=1$ primary 
resonance sits at a higher energy than all of the other resonances.  
As the strength of the driving field is increased the resonances 
overlap and the dynamics in the region of overlap becomes chaotic.  At 
very high field strengths all higher order resonances have been 
destroyed and only the $N=1$ resonance remains.  As the field strength 
is increased still further, even this $N=1$ resonance begins to 
disappear.  However, since there are no resonances at higher energies 
the region of chaos remains bounded from above by regular motion.  
This bounded chaos can be seen clearly in strobe plots of the 
classical motion of this system.  Figure\ \ref{strobe} shows strobe 
plots of this system for $\epsilon = 175$ and $750$.  The coordinates 
for these plots are the action-angle variables of the undriven square 
well, defined by $J=2|p|/\pi$ and $\theta=\pm \pi(x+1)/2$ where all 
quantities are dimensionless because of the above scaling.

\subsection{Quantum Dynamics and Floquet Theory}

The Schr\"{o}dinger equation for a driven particle in an infinite
square well is given by
\begin{equation}
i \hbar \frac{\partial}{\partial \tilde{t}} |\psi(\tilde{t}) \rangle = 
-\frac{\hbar^{2}}{2m} \frac{\partial^{2}}{\partial \tilde{x}^{2}} + 
\tilde{\epsilon} \tilde{x} \cos(\tilde{\omega_{0}} \tilde{t})
\end{equation}
where $|\tilde{x}| \leq a$.  All parameters are defined as in the 
classical Hamiltonian above.  The transformation to dimensionless 
coordinates is identical to that used for the classical Hamiltonian, 
except that instead of scaling $\tilde{p}$ we must scale $\hbar$.  The 
resulting dimensionless equation is
\begin{equation}
i \kappa \frac{\partial}{\partial t} | \psi(t) \rangle = \left( - 
\kappa^{2} \frac{\partial^{2}}{\partial x^{2}} + \epsilon x 
\cos(\omega_{0} t) \right) | \psi(t) \rangle
\end{equation}
where energy is measured in units of $c$ (as in the classical case) and 
$\kappa =\hbar/(a\sqrt{2mc})$.  Note that in the 
quantum system there are three parameters: $\epsilon$, $\omega_{0}$, and 
$\kappa$.  Only two of these parameters are independent, so the full 
dynamics of this system can be studied by varying two parameters.  
In this paper we will only vary one of the parameters, $\epsilon$, and 
we set $\kappa=1$ and $\omega_{0}=80$ as above.  The effect of 
varying $\kappa$ is left for future study.

This quantum system is periodic in time and can be described in terms of
 Floquet eigenstates \cite{shirley,sambe},
which are simply eigenstates of the one-period time evolution
operator.  Since the time evolution operator is unitary for this
system (there is no ionization) all of the eigenvalues of the operator
have modulus 1.  Thus the Floquet states satisfy
\begin{equation}
\hat{U}(T)\,|\Omega_{\alpha}\rangle= e^{-i\Omega_{\alpha} T/\kappa}\,
|\Omega_{\alpha}\rangle,
\end{equation}
where $\Omega_{\alpha}$ is the Floquet eigenvalue (or quasienergy) and 
$T$ is the period of the driving field.  It is apparent from the above 
equation that the quasienergy is only defined modulo 
$\omega_{0}=2\pi/T$ (with $\kappa=1$).

The Floquet states can be computed numerically by first calculating
the matrix for the operator $\hat{U}(T)$ in a basis of unperturbed
square well eigenstates.  A numerical diagonalization of this matrix
produces the eigenvalues and eigenvectors (in the unperturbed basis).
Since there is little coupling between
states in the regular region and states in the region of bounded
chaos, a basis that extends into the regular region will give an
accurate description of all states associated with the chaotic region 
\cite{breuer1}.
In our calculations we use a basis of 80 eigenstates.  With $\kappa=1$
this basis extends far into the regular region for all values of
$\epsilon$ that we will consider here.

\section{Avoided Crossings}
\label{crossings}

We wish to study the quantum dynamics that takes place near an avoided 
crossing (AC) in the spectrum of quasienergies for this system.  Our first 
step then is to locate some avoided crossings.  Figure \ref{qnrg} 
shows the (mod $\omega_{0}$) spectrum of the 40 lowest quasienergies 
as a function of $\epsilon$.  As mentioned in Section \ref{intro}, 
there is a close connection between the onset of ACs and 
chaos in the classical system.  Figure \ref{qnrg} shows that avoided 
crossings begin to appear between $\epsilon=100$ and $\epsilon=200$, 
and by $\epsilon=800$ the spectrum is dominated by avoided 
crossings.  This compares well with the growth of the chaotic region 
in the classical phase space between these two values of $\epsilon$ 
(see Fig.\ \ref{strobe}).

It is important to note that there are many places, particularly at 
small values of $\epsilon$, where the quasienergy curves actually 
cross.  This happens when the two states associated with the curves 
belong to different (uncoupled) sectors of the Hilbert space and 
transitions between these states are forbidden 
\cite{wigner,berry,breuer2}.  By different sectors of Hilbert space we 
mean that the two states belong to different blocks of a 
block-diagonal Hamiltonian, indicating that they belong to different 
symmetry groups.  When $\epsilon$ is small these ``apparent 
crossings'' are quite common, but at large $\epsilon$ there are few 
apparent crossings.  The reason for this is that the spread of chaos 
is due to the breaking of the same symmetries that prevent coupling of 
certain states.

There are many strong ties between avoided crossings and chaos,
beyond the observed increase of ACs as the classical
system becomes chaotic.  Successive avoided crossings are responsible for the
transition to a random-matrix distribution \cite{wang}, a property which
has long
been associated with chaos.  In fact, there is a strong correlation
between the overlap of these successive ACs and the
fraction of the classical phase space which is chaotic \cite{yang}.
Other studies have shown that ACs occur only when the
two states can tunnel through any Kolmogorov-'Arnold-Moser (KAM) tori that
lie between them in the phase space \cite{henkel}.  This means that
the states involved in avoided crossings tend to lie in regions of
the phase space where the KAM tori have been strongly distorted or
destroyed altogether \cite{sherwin1}.

By studying Fig.  \ref{qnrg} we can formulate a general picture of
what is happening to the quasienergy curves of this system as
$\epsilon$ is increased.  At low $\epsilon$ there are no avoided
crossings and most of the quasienergy curves maintain a constant
slope.  The only curvature here is in the set of curves that look like
the characteristic curves of the Mathieu equation \cite{handbook}.
The states associated with these curves are becoming trapped in the
pendulum-like $N=1$ primary resonance that forms as $\epsilon$ increases
from 0.  The curve that looks like the ground state of the Mathieu
equation is connected to the n=16 square well state at $\epsilon=0$.
Note that $J=16$ is the exact position of the $N=1$ resonance in the
strobe plots of Fig.  \ref{strobe}.  At higher values of $\epsilon$
ACs begin to appear.  At the avoided crossing itself
there is a significant change in the slope of the quasienergy curves,
but at these moderate values of $\epsilon$ the AC seems
to result only in an exchange of slopes between the two curves.  At
the highest values of $\epsilon$ shown in Fig.  \ref{qnrg} the avoided
crossings result in dramatic changes in the slopes of the quasienergy
curves, not just an exchange of slope.  One may also note that the
Mathieu curves are no longer identifiable at these values of
$\epsilon$.  At these high values of $\epsilon$ the $N=1$ resonance has
become highly distorted and is beginning to disappear into the chaotic
sea (see Fig.  \ref{strobe}b).  Thus we see that there is a strong
connection between changes in the classical phase space and changes in
the quasienergy spectrum.

\section{Changes in Floquet State Structure at an Avoided Crossing}
\label{structure}

This connection between chaos and quasienergy curves is interesting, 
but it is not entirely clear.  For one thing, only the connection 
between very large changes in the classical phase space and 
correspondingly large changes in the quasienergy spectrum has been 
established.  We would like to study the changes that take place at a 
single avoided crossing.  Additionally, we would like to see changes 
in the structure of the Floquet states, rather than changes in the 
quasienergy curves.  To see these structural changes we must construct 
a quantum mechanical ``phase space''.  We can then monitor the 
changes in the ``phase space'' as the value of $\epsilon$ moves 
through an avoided crossing.

\subsection{Visualization of Quantum Phase Space}

Visualization of the quantum mechanical ``phase space'' requires the
construction of a phase space probability density for the various
quantum eigenstates.  The uncertainty principle prevents the
construction of a true phase space probability density in the
classical sense, but it is possible to construct a quasiprobability
density that is positive definite and gives a coarse-grained picture
of the distribution of the quantum state in phase space.  This
quasiprobability density is the Husimi distribution
\cite{husimi,takahashi}.  To construct the Husimi distribution of a
given quantum state one simply calculates the overlap between the
given state and a coherent state centered on a particular point
$(x_{0},p_{0})$ in phase space.  The wavefunction of the coherent
state in configuration space is \cite{crespi}
\begin{equation}
\langle x|x_{0},p_{0}\rangle=
\left(\frac{1}{\sigma^2\pi}\right)^{1/4}\exp
\left(-\frac{(x-x_{0})^2}{2\sigma^2}+
\frac{ip_{0}(x-x_{0})}{\hbar}\right),
\end{equation}
where $\sigma$ is a squeezing parameter that determines the relative
widths of the coherent state in the $x$ and $p$ directions.
Calculating $\left| \langle \psi|x_{0},p_{0}\rangle \right|^{2}$ for a
grid of phase space points will produce a quasiprobability
distribution that can be easily viewed as a contour plot.  Figures
\ref{hus1} and \ref{hus2} show Husimi distributions for several
Floquet states in this system.

Now we can use the Husimi distribution to visualize changes in
the structure of the quantum phase space that take place at an
AC.  We will examine the Husimi distributions of
Floquet states at values of $\epsilon$ slightly less than, slightly
greater than, and at the value at which that state undergoes an
AC.  This will allow us to determine what changes occur
at the AC and to what extent these changes survive at
higher values of $\epsilon$.

A close inspection of Fig.  \ref{qnrg} reveals that not all avoided
crossings look the same.  As discussed in Sec.  \ref{crossings} there
are crossings where the curves simply exchange slopes and crossings
where the slopes change.  We will refer to the crossings that exchange
slopes as sharp ACs.  The states involved in such an AC are weakly
coupled and usually lie in different regions of the phase space
(inside a resonance and in the chaotic region, for instance).  The
other type, broad ACs, involve strongly coupled states that usually
reside in the same region of phase space.  In fact, most broad ACs
occur between states that are associated with the region of chaos.

\subsection{Sharp Crossing}
\label{sharp}

Figure \ref{qnrgsharp} is a detail from Fig. \ref{qnrg} that focuses on
a sharp avoided crossing at $\epsilon \approx 175$.  The two curves 
that participate in the avoided crossing are labeled A and B.  In Fig.
\ref{hus1} we show the Husimi distributions of the two Floquet states
at $\epsilon = 170$, $175.5$, and $180$.  At $\epsilon=170$ we can see
that state A is contained within the n=1 primary resonance
(see Fig. \ref{strobe}a for a picture of the classical dynamics near
this field strength) while state B lies in the low-energy chaotic
region.  At $\epsilon=175.5$, the center of the avoided crossing, the
Husimi distributions for both states appear to be mixtures of the
states shown for $\epsilon=170$.  Clearly the AC has a dramatic
impact on the structure of these states at this particular value of
$\epsilon$.  However, the Husimi distributions at $\epsilon=180$ show
that these changes do not persist at higher field strengths.  The two
states simply exchange their structure, so that the net effect is a
relabeling of the Floquet states.  So away from the AC itself the
overall structure of the quantum phase space is unchanged.

These dynamics can be understood quite well using a two-level
approach \cite{jie}.  These sharp ACs really only involve the two
Floquet states whose curves nearly come together.  The rest of the
Hilbert space has very little influence on the dynamics of these two
states.  Without any contribution to the dynamics from other states,
these two levels can only exchange their structure.  These types of
crossings play an important role in tunneling, since a state
originally confined to a resonance has become a chaotic state after
the AC \cite{latka1}, but they do not play a significant role in
altering the structure of the quantum phase space.

\subsection{Broad Crossing}
\label{broad}

In a broad AC several states make significant contributions to the 
dynamics.  A particularly striking example of this is shown in Fig.  
\ref{qnrgbroad}, where two ACs (at $\epsilon \approx 750$ and 
$\epsilon \approx 765$) actually overlap.  Here there are three states 
(labeled C, D, and E) that are strongly influencing each others' 
dynamics.  There is no simple exchange of slopes between the curves in 
this AC. Hence, we might expect to find more interesting (and 
permanent) structural changes in this crossing than in the previously 
studied one.  Figure  \ref{hus2} shows the effect of the avoided 
crossing on the structure of the three states that are involved.  As 
in the sharp crossing there is a mixing of structures for values of 
$\epsilon$ that lie within the crossing region.  However, in this case 
the changes do not disappear when we look at larger values of field 
strength.  With this type of AC the states are not simply 
``relabeled'', but undergo actual changes in their phase space 
structure.  Particularly striking is the difference between Figs.  
\ref{hus2}c and \ref{hus2}h.  These states would be identical if there 
was a complete exchange of structure as seen in the sharp AC. Instead, 
before the crossing state E is localized at very low energies, but 
after the crossing state D (which has exchanged most of its structure 
with state E) has spread into the higher energy portion of the region 
of chaos.  The avoided crossing has delocalized this particular 
Floquet state.  We find this result to be quite general, that broad 
ACs lead to permanent changes in the structure of Floquet states that 
tend to delocalize the states.  Of course, since the region of chaos 
is bounded the states can only delocalize until they reach the 
boundaries of the chaos.  At extremely high values of $\epsilon$, 
where nearly every ``chaotic'' state has undergone many broad avoided 
crossings, we find that all of these states are delocalized and fill 
the chaotic region \cite{chism}.

\section{Effect of Structural Changes on Radiation Spectra}
\label{spectra}

Now that we have seen how avoided crossings can affect the structure 
of Floquet states we would now like to see how they effect an 
experimentally observable quantity, namely the radiation spectrum.  In 
a prior work we found that the generation of high harmonics increased 
as the system, initially in a single Floquet state, passed through the 
avoided crossing \cite{chism}.  However, this effect was caused by 
population transfer from the original Floquet state to the other state 
as the field strength was increased.  This population transfer creates 
a superposition of two Floquet states that displays increased 
radiation at high frequencies (albeit shifted away from the harmonic 
frequencies).  In this study we would like to focus on changes in the 
radiation spectrum that are caused by the changing structure of a 
single Floquet state.  For this reason we will calculate radiation 
spectra by starting the system in a given Floquet state and 
maintaining a constant field strength for 128 cycles of the driving 
field.  The expectation value of the acceleration for the state is 
calculated during this time interval.  We then calculated the Fourier 
transform, $\xi(\omega)$, of this acceleration time series.  The 
square modulus of the Fourier transform gives us the radiation 
spectrum.  Because the Floquet states are periodic with period 
$2\pi/\omega_{0}$ it is not truly necessary to calculate for 128 
cycles of the driving field.  However, this long integration time 
exposes numerical errors that might be hidden in a shorter 
calculation.

To study the effect of structural changes on radiation spectra we 
cannot simply study the spectrum of a single Floquet state for various 
values of $\epsilon$.  Avoided crossings cause states to exchange 
structure, effectively relabeling the states.  If true structural 
changes are to be distinguished from simple relabeling, one must 
account for this relabeling when comparing different spectra.  At the 
midpoint of the avoided crossing this is nearly impossible to do, 
since the relabeling has not fully taken effect.  For values of 
$\epsilon$ that are beyond the AC, the relabeling can easily be taken 
into account.  Our procedure in the following is to calculate spectra 
for a state before the AC, the same state (on the same curve) at the 
midpoint of the AC, and the relabeled state (now on a different 
quasienergy curve) after the AC. This separates the changes in 
radiation spectrum that occur because of structural change in the 
Floquet state from the apparent changes that occur because the states 
have been relabeled.

\subsection{Sharp Crossing}

We first calculate radiation spectra for the states whose Husimi 
distributions are shown in Fig.  \ref{hus1}(a,c, and f).  The first 
two states are associated with the curve A in Fig.\ \ref{qnrgsharp}, 
while the third is associated with curve B. By changing from A to B 
after the avoided crossing we can account for the effects of 
relabeling as described above.  The spectra are shown in Fig.  
\ref{spec1}.  Between $\epsilon=170$ and $\epsilon=175.5$ there is a 
significant increase in the radiation at the highest harmonics 
(11-19).  However, these harmonics have decreased at $\epsilon=180$.  
This increase and subsequent reduction is easier to see in Fig.  
\ref{spec2}, which shows the differences between the spectrum at 
$\epsilon=170$ and the spectra at $\epsilon=175.5$ and $180$.  This 
temporary increase in high harmonic generation is exactly what we 
expect from the temporary changes in the phase space structure seen in 
Fig.  \ref{hus1}.  At the midpoint of the AC both states are a mixture 
of the two states at $\epsilon=170$ and both are spread over a wider 
region of phase space.  This leads to an increase in harmonic 
generation at this value of $\epsilon$.  After the avoided crossing, 
however, these structural changes disappear (with the exception of the 
relabeling) and the harmonic generation subsides.  Thus, sharp ACs 
only affect the radiation spectrum of a Floquet state for field 
strengths that lie within the crossing.

\subsection{Broad Crossing}

Now we investigate the radiation spectra for the states shown in Fig.  
\ref{hus2}(c, f, and h).  Again we switch from state E to state D for 
values of $\epsilon$ that are beyond the avoided crossing, to account 
for the relabeling that takes place.  Fig.  \ref{spec3} shows the 
spectra for these states.  There is a steady increase in the radiation 
at the highest harmonics (31-45) as $\epsilon$ is increased.  This is more 
easily seen in Fig.  \ref{spec4} which shows the differences between 
the spectra.  The changes in the spectra are quite complicated, but 
there is clearly no reversal in the increase of high harmonics as 
was observed in the sharp crossing.  This broad AC permanently 
increases the high harmonic generation.  This is closely tied 
to the delocalization that is observed in Fig.  \ref{hus2}, since the 
generation of high harmonics depends on the number of energy levels 
over which the Floquet state is spread \cite{chism}.

\section{Conclusion}
\label{conclusion}

We have seen here that avoided crossings can dramatically alter the 
phase space structure of Floquet states in a driven quantum system.  
Sharp ACs, involving states in separate sectors of the Hilbert space, 
produce structural changes that do not survive at higher field 
strengths.  The only effect that remains at higher field strengths is 
a relabeling of the Floquet states.  These crossings increase the HHG 
from a Floquet state at the crossing itself, but they do not lead to 
increased HHG at higher field strengths.  Broad crossings, however, 
usually involve states in the same sector of the Hilbert space and 
often include effects from several states at once.  These crossings 
can create persistent changes in the structure of the Floquet states 
and in the radiation spectra that these states produce.  At field 
strengths beyond the avoided crossing the states will be less 
localized and the spectra will show stronger radiation at high 
harmonics.

The data on harmonic generation is interesting, as high harmonic
generation has gained a lot of attention in recent years.  But perhaps
even more intriguing are the results that show how avoided crossings
delocalize Floquet states.  This has bearing on the important problem
of the stabilization of atoms in intense laser fields.  Theoretical
studies have shown that this stabilization may result from the
electron occupying a Floquet state that is localized in the phase
space \cite{sundaram1}.  This localization prevents the electron from
reaching the continuum and thus decreases the ionization rate.  To
better understand stabilization one must understand how localized
Floquet states are created and destroyed.  Our work indicates that
avoided crossings play a key role in this process.  It must be noted,
however, that the model used in this paper does not allow ionization.
To study stabilization this work must be extended to open quantum
systems.

Although the primary motivation for this study was theoretical, these 
results may be experimentally observable.  The infinite square well 
serves as a simple model for recent experiments on 
electron confinement in GaAs/AlGaAs quantum 
wells \cite{sherwin1,sherwin2}.  These experiments confine electrons 
in wells that vary in width from 50 to 1000 \AA \ and in depth from 
200 to 300 meV.  A well with a width of only 50 \AA \ and a depth of 
300 meV contains only a few bound states and therefore cannot be 
expected to produce the effects seen in this study. However, a well 
with a depth of 300 meV and a width of 600 \AA \ contains about 50 bound 
states.  The dynamics of such a well, driven by a far-infrared laser at low 
intensity, should  be similar to the dynamics of our model.  The 
parameters used in this paper correspond to a laser with a wavelength of 400 
$\mu$m and intensity of $10^{5} \mathrm{W/cm}^{2}$ striking a 600 \AA 
\ well with a depth of 300 meV.  These parameters are well within the 
range accessible by recent experiments.

\acknowledgments

The authors wish to thank the Welch Foundation Grant No. F-1051 and
DOE Contract No. DE-FG03-94ER14465 for partial support of this work.
We also thank NPACI and the University of Texas at Austin High
Performance
Computing Center for use of their computer facilities.

\appendix

\begin{figure}
\caption{Strobe plots of the classical dynamics of the driven square 
well at two field strengths.  $J$ and $\theta$ are the dimensionless 
action-angle variables for the undriven square well.  In (a) 
$\epsilon=174$ and the primary $N=1$ resonance is very prominent 
within the region of bounded chaos.  In (b) $\epsilon=780$ and the 
$N=1$ resonance is distorted and occupies a smaller region of phase 
space than at $\epsilon=174$.  The chaotic region is much larger at 
this higher field strength.}
\label{strobe}

\end{figure}

\begin{figure}
\caption{Spectrum of quasienergies for the driven square well as a function of 
field strength.  The first avoided crossings appear between 
$\epsilon=100$ and $\epsilon=200$.  As $\epsilon$ is increased the 
number of avoided crossings increases rapidly so that by 
$\epsilon=800$ avoided crossings dominate the spectrum.  Both 
$\Omega_{\alpha}$ and $\epsilon$ are dimensionless quantities.}
\label{qnrg}

\end{figure}

\begin{figure}
\caption{Detail from Figure\ \ref{qnrg} showing a sharp avoided 
crossing near $\epsilon=175.5$.  The quasienergy curves involved in the 
avoided crossing are labeled A and B.} 
\label{qnrgsharp}

\end{figure}

\begin{figure}
\caption{Husimi distributions of the Floquet states involved in the 
avoided crossing shown in Figure\ \ref{qnrgsharp}.  The labels A and B 
indicate the quasienergy curve in Fig.\ \ref{qnrgsharp} with which the 
state is associated.  The $\epsilon$ values indicate the field 
strength at which the Floquet state was calculated.  At 
$\epsilon=175.5$ (the center of the avoided crossing) both Floquet 
states are mixtures of the two states at $\epsilon=170$.  By 
$\epsilon=180$ A and B have exchanged their structure 
completely.  Note that the coordinates for all of the Husimi plots are the 
dimensionless action-angle variables used in Fig.\ \ref{strobe}.}
\label{hus1}

\end{figure}

\begin{figure}
\caption{Detail of Figure\ \ref{qnrg} showing a pair of broad avoided crossings 
near $\epsilon=760$.  The three quasienergy curves that are involved 
in the avoided crossing are labeled C, D, and E.}
\label{qnrgbroad}

\end{figure}

\begin{figure}
\caption{Husimi distributions of the three Floquet states involved in 
the crossings shown in Figure\ \ref{qnrgbroad}.  The labels C, D, and 
E indicate the quasienergy curve in Fig.\ \ref{qnrgbroad} with which 
the state is associated.  At $\epsilon=760$ all three states are 
mixtures of the states at $\epsilon=740$.  By $\epsilon=780$ the 
states have exchanged most of their structure, but there are significant 
differences from the $\epsilon=740$ states.  In particular,
we expect the states (c) and (h) to have similar structure but 
instead find that (h) is much less localized than (c).}
\label{hus2}

\end{figure}

\begin{figure}
\caption{Radiation spectra generated by the Floquet states shown in 
Figure\ \ref{hus1}(a,c,e).  There is a significant increase in 
harmonic generation from $\epsilon=170$ (a) to $\epsilon=175.5$ (b).  
However, by $\epsilon=180$ (c) this increase has disappeared.  All 
spectra have been normalized so that they have the same power at the 
fundamental frequency $\omega_{0}$.}
\label{spec1}

\end{figure}

\begin{figure}
\caption{Differences between spectra shown in Figure\ \ref{spec1}.  
The difference between Fig.\ \ref{spec1}b and Fig.\ \ref{spec1}a is 
shown in (a).  The difference between \ref{spec1}c and \ref{spec1}a 
is shown in (b).  The increase and subsequent decrease in harmonic 
generation is apparent.}
\label{spec2}

\end{figure}

\begin{figure}
\caption{Radiation spectra generated by the Floquet states shown in 
Figure\ \ref{hus2}(c,f,h).  There is a significant increase in 
harmonic generation from $\epsilon=740$ (a) to $\epsilon=760$ (b) and 
by $\epsilon=780$ the harmonic generation has increased even further.  All 
spectra have been normalized so that they have the same power at the 
fundamental frequency $\omega$.}
\label{spec3}

\end{figure}

\begin{figure}
\caption{Differences between spectra shown in Figure\ \ref{spec2}.  
The difference between Fig.\ \ref{spec2}b and Fig.\ \ref{spec2}a is 
shown in (a).  The difference between \ref{spec2}c and \ref{spec2}a 
is shown in (b).  The increase in harmonic generation seen in (a) 
clearly persists in (b).}
\label{spec4}

\end{figure}

\end{document}